\documentclass[screen,authorversion,nonacm]{acmart}
\pdfoutput=1
\usepackage{url}
\usepackage{fancyhdr}

\calclayout

\AtBeginDocument{%
    \addtolength{\footskip}{2.0\baselineskip}%
    \fancyfoot[L]{
      \hrule
      \vspace*{0.3\baselineskip}
      © Irene Weber 2021. This is the author's version of the accepted version of the paper. It is posted here for
      your personal use. Not for redistribution. The final published version of the paper is
      in CUI'21 {3rd Conference on Conversational User Interfaces}, {July 27--29, 2021}, {Bilbao (online), Spain}.
      It can be accessed at \url{https://doi.org/10.1145/3469595.3469632}
    }}

\begin{document}

% \title[Low-code from frontend to backend]{\Large Low-code from frontend to backend: Connecting conversational user
% interfaces to backend services via a low-code IoT platform
% \vspace*{\baselineskip}}
\title[Connecting conversational user
interfaces to backend services via a low-code IoT platform]{\Huge Low-code from frontend to backend: Connecting conversational user
interfaces to backend services via a low-code IoT platform
\vspace*{\baselineskip}}

\author{Irene Weber}
\email{irene.weber@hs-kempten.de}
\orcid{0000-0003-2743-1698}
\affiliation{%
  \institution{Kempten University of Applied Sciences}
  \city{Kempten}
  \country{Germany}
}

\begin{abstract}
    Current chatbot development platforms and frameworks facilitate setting
    up the language and dialog part of chatbots, while connecting it to
    backend services and business functions requires substantial manual
    coding effort and programming skills. This paper proposes an approach to
    overcome this situation. It proposes an architecture with a chatbot as
    frontend using an IoT (Internet of Things) platform as a middleware for
    connections to backend services. Specifically, it elaborates and
    demonstrates how to combine a chatbot developed on the open source
    development platform Rasa with the open source platform Node-RED,
    allowing low-code or no-code development of a transactional
    conversational user interface from frontend to backend.
\end{abstract}





\maketitle

\vspace{2\baselineskip}

% \subsubsection*{Authors address}
% Irene Weber, irene.weber@hs-kempten.de, Kempten University of Applied Sciences, Kempten, Germany

\subsection*{Versions}
© Irene Weber 2021. This is the author's version of the accepted version of the paper. It is posted here for
your personal use. Not for redistribution. The final published version of the paper is 
in CUI'21 {3rd Conference on Conversational User Interfaces}, {July 27--29, 2021}, {Bilbao (online), Spain}.
It can be accessed at \url{https://doi.org/10.1145/3469595.3469632}

\thispagestyle{plain}
\section{Introduction}

Chatbots offer a means to use services and control smart home devices
connected to the web. They can also serve as conversational user
interfaces to application systems in other contexts, including
professional work. Chatbots that accomplish tasks for users are termed
transactional chatbots, as opposed to conversational chatbots that
engage in dialogs with users mainly for social reasons or entertainment
\cite{mctear_conversation_2018}.This paper addresses transactional chatbots serving as
conversational user interfaces.

Compared to graphical user interfaces, conversational user interfaces
can bring several advantages: They provide a more natural,
user-controlled, and flexible way to interact with computer systems than
graphical user interfaces. They make working with computer systems
easier, especially when users have to complete tasks only occasionally.
As a result, chatbots have become popular as user interfaces for
customer self-service functions and the like.

Conversational user interfaces also have great potential in a
professional environment as they can help employees complete non-routine
tasks more smoothly and efficiently, such as filling request forms,
retrieving information from knowledge bases, adding new entries to
knowledge bases, filing reports, and more. Conversational user
interfaces can serve as a single and unified entry point to multiple
system functions and services, eliminating the need for users to become
familiar with various application systems and software tools and their
respective page, tab, and menu structures.

In order to perform tasks on behalf of users, a transactional
conversational user interface, serving as a frontend, needs to access
services, devices, databases, and application systems as backend.
Consequently, the connectivity of the conversational user interface's
language-processing components to external systems and services is
critical.

The technical aspects of the overall architecture of a system with a
conversational user interface have received rather little attention in
research. Attention may increase recently, as it may indicate
interesting research opportunities. For instance, Baez et al \cite{baez_chatbot_2020}
conducted a literature search and reviewed papers that address
architectural aspects of integrating chatbots with existing external
systems. They identified seven integration patterns. Their goal is to
stimulate research on development aids that specifically support the
development of bots' language capabilities for a given integration
pattern. They also note that they have found little work on integrating
a conversational user interface for specific use cases (namely,
\emph{API caller} and \emph{business process interface}), despite the
great practical potential that these use cases promise. Rough and Cowan
\cite{rough_dont_2020} surveyed current chatbot development platforms and frameworks.
They found that the platforms and frameworks make it easy to set up the
voice and dialog portion of a chatbot, while connecting to backend
services and business functions requires significant manual coding
effort and programming skills.

This paper proposes a system architecture with a chatbot as front\-end
using an IoT (Internet of Things) platform as a middleware for
connections to backend services. Specifically, it elaborates and
demonstrates how to combine a chatbot developed on the open source
development platform Rasa with the open source platform Node-RED,
allowing low-code or no-code development of a transactional
conversational user interface from frontend to backend.

\section{Architecture of Conversational User Interfaces}

\subsection{Standard architecture of a conversational user
interface}

\begin{figure}[t]
  \centering
  \includegraphics[width=0.8\linewidth]{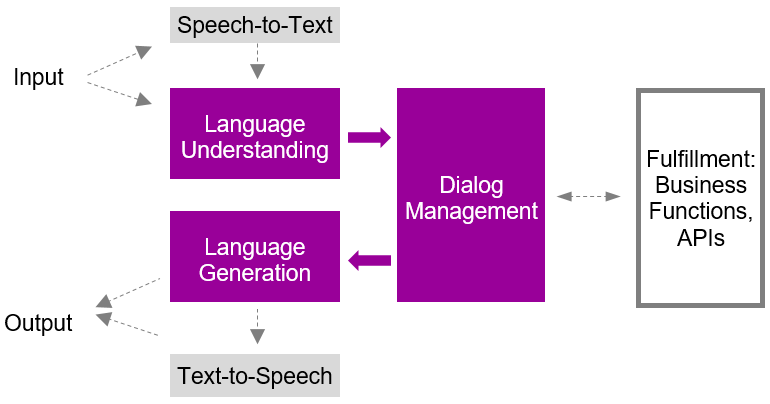}
  \caption{Typical architecture of a conversational interface, adapted from
  \cite{mctear_conversation_2018,rough_dont_2020}}\label{fig-cui-arch}
  \Description{A block diagram with blocks labeled Input, Speech-to-Text, Language Understanding, 
  Dialog Management, Fulfillment, Language Generation, Text-to-Speech, and Output, 
  and connecting arrows }
\end{figure}

The main components of a chatbot are Language Understanding, Dialog
Management, and Language Generation. The chatbot may accept typed input
or spoken input and produce written or speech output. When conversing by
speech, the architecture has to provide for text-to-speech- and
speech-to-text transformation. Figure~\ref{fig-cui-arch} shows a typical architecture of
a conversational interface \cite{mctear_conversation_2018,rough_dont_2020}.

The Language Understanding component analyzes language input and maps it
to intents and entities. The Language Generation component composes the
output of the chatbot. The Dialog Management component controls the
chatbot's behavior. It also holds the slots that implement the chatbot's
memory where it stores entities and the state of the dialog.

The Dialog Management component determines how the chatbot acts upon
user inputs, given the recognized intents, entities, slot values, and
dialog history so far. In any case, the chatbot eventually provides some
output to the user in order to continue the dialog. Before doing so, it
may access backend systems to perform a task or retrieve information and
include the outcome in its output to the user. The component performing
tasks and calling external services constitutes the Fulfillment part of
the chatbot.

As \cite{rough_dont_2020} have stated, current chatbot development platforms provide
good support and tools for developing the Language Understanding, Dialog
Management, and Language Generation components, while implementing the
Fulfillment part and its interfaces to backend systems requires manual
programming skills and effort. This complicates building a
conversational user interface, increases cost and development time, and
prevents end users from setting up a CUI as their personal
conversational assistant. In addition, this is a significant
disadvantage for training and hands-on classroom exercises when teaching
on conversational user interfaces. This work proposes an approach to
alleviate this problem for the chatbot development platform Rasa
\cite{rasa_technologies_inc_open_2020}.

\subsection{Proposed architecture for Rasa CUIs with Node-RED}

In the architecture of a Rasa chatbot, a so-called action server handles
fulfillment. The action server runs as a standalone server and bundles
and manages calls to external APIs. The Rasa platform provides a Python
SDK for developing an action server executing custom actions. While the
SDK facilitates the development, Python coding is still necessary.

\begin{figure}[t]
    \centering
    \includegraphics[width=0.8\linewidth]{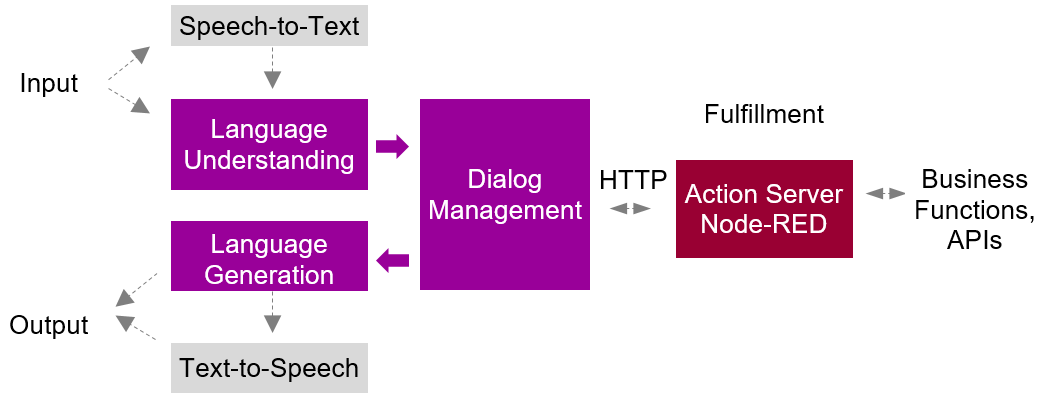}

    \caption{Architecture of a conversational user interface based on Rasa}\label{fig-rasa-arch}
    \Description{A block diagram with blocks labeled Input, Speech-to-Text, Language Understanding, 
    Dialog Management, Fulfillment, Business Funtions, Language Generation, Text-to-Speech, and Output, 
    and connecting arrows. 
    The Fulfilment block shows the label ``Action server, Node-RED''}
  \end{figure}

The approach proposed here replaces the Python-coded action server by an
action server hosted on Node-RED. Figure~\ref{fig-rasa-arch} depicts the resulting
architecture.

\section{Implementing the Rasa Action Server HTTP API on Node-RED}

This section gives an example on how to implement an action server, i.e.
a fulfillment component, for a Rasa chatbot by means of visual
programming on the Node-RED platform. It introduces the basics of
Node-RED, explains different types of nodes that are useful for building
an action server, and shows a complete example flow.

\subsection{ Node-RED Basics}

Node-RED is a low code platform running on Node.js. It is especially
popular for connecting IoT devices and developing, e.g., home automation
applications. It brings a browser-based GUI with a visual flow editor
and a palette of nodes. Nodes are drag-and-dropped from the palette onto
the flow editor and interconnected to flows. When triggered, a flow
transports a message object, typically coded in JSON, from node to node,
with the nodes augmenting or transforming the message object and,
optionally, performing additional activities, such as calls to further
services or APIs. Information extracted from the message object can
parameterize the API calls. Different types of events can trigger a
flow, one of them being an incoming HTTP request. It is also possible to
trigger a flow manually for development and testing.

The Node-RED palette holds a basic set of standard nodes. Different
types of nodes implement conditional branching or loops, manipulate
message objects, communicate with external systems as a server or as a
client, catch and handle errors, or display debugging information.

Besides the standard nodes, Node-RED can import and use custom nodes.
The Node-RED ecosystem offers a broad variety of community-contributed
nodes, including nodes that, e.g., connect to databases, access web
APIs, write to spreadsheets, control devices, and more. If needed,
experienced programmers may implement own custom nodes.

\subsection{Custom nodes implementing the Rasa action server HTTP API}

An action server for a Rasa chatbot must implement the HTTP API of Rasa
action servers as specified in \cite{rasa_technologies_gmbh_rasa_2021}. The author provides a collection
of specialized nodes, node-red-contrib-rasa-actionserver \cite{weberi_weberinode-red-contrib-rasa-actionserver_2021},
abbreviated as \emph{rasaas,} that facilitates building the action
server for Rasa on Node-RED. The \emph{rasaas} nodes are publicly
available for download. Node-RED allows searching and installing these
and other custom nodes from its graphical user interface. After
installation, the nodes become available in the Node-RED palette.

\subsection{Example application}

\begin{figure}[ht]
    \centering 
    \includegraphics[width=0.8\linewidth]{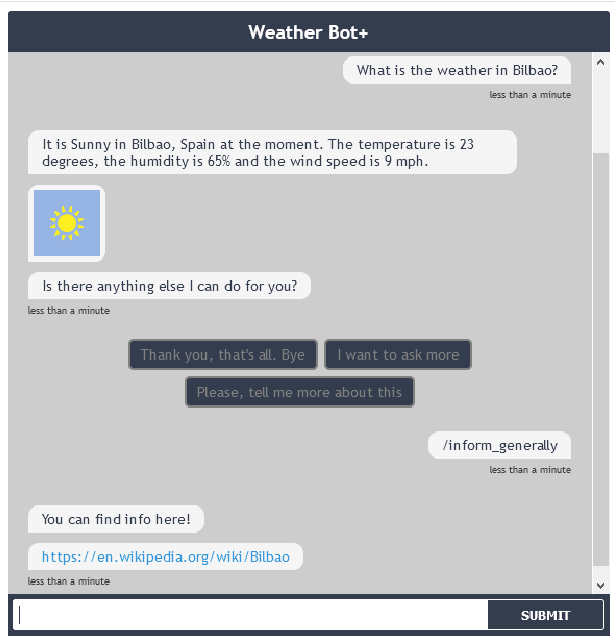}
    \caption{Sample dialog, using the chatroom of \cite{scalableminds_github_2020}}\label{fig-chatroom}
    \Description{A chatbot input/output box with some text, an icon, buttons, and a hyperlink. }
  \end{figure}

The application is a simple chatbot that provides information about
geographical locations, such as cities or regions. Specifically, it can
provide information about the current weather or provide a link to the
Wikipedia page of a location named by the user. This standard sample
application is an extended version of \cite{petraityte_justinapetrweatherbot_tutorial_2020}. It is sufficient to
demonstrate the approach, since it accesses two external APIs
(weatherstack.com and Wikipedia OpenSearch). For better understanding, a
sample dialog with the bot is shown in Figure~\ref{fig-chatroom}.

\subsection{Building the action server flow}

\subsubsection{HTTP Endpoint}
Rasa dialog management sends a HTTP request to the action server
carrying information about the state of dialog coded in JSON. The basic,
most relevant pieces of information are the name of the specific action
that the action server has to perform, and slots with their current
values.

Enabling communication between the chatbot's Dialog Management and its
Node-RED action server is straightforward, as the sole requirement for a
custom action server is to provide an HTTP endpoint conforming to the
HTTP API specification of Rasa action servers \cite{rasa_technologies_gmbh_rasa_2021}. Node-RED provides
a pair of nodes that implement an HTTP endpoint, the \emph{http in} and
\emph{http response} nodes. This pair of nodes represent start and end
of a Node-RED flow implementing an action server for RASA.

\begin{figure*}[t]
    \centering
    \includegraphics[width=\linewidth]{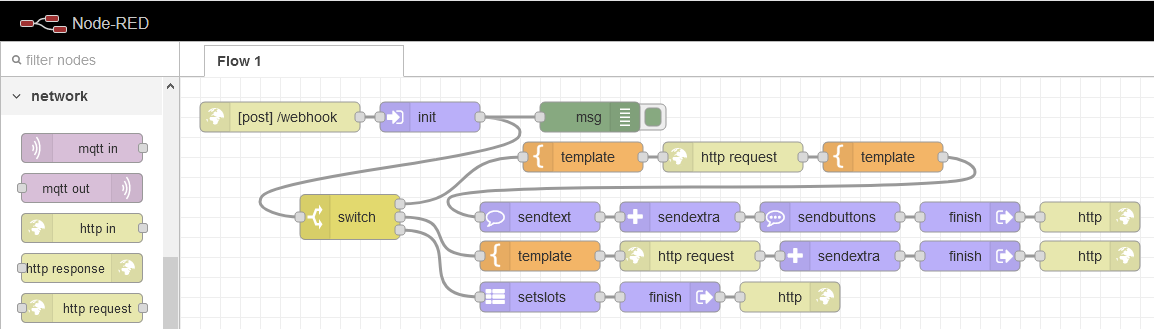}
    \caption{Node-RED Flow editor with an example flow implementing an action server, i.e., fulfillment component, 
    for a Rasa conversational user interface. 
    To the left, the network section of the Node-RED palette is partially visible. }\label{fig-sample-flow}
    \Description{A graphical user interface with a structure of labelled nodes that are connected to a flow with several branches }
  \end{figure*}

The \emph{http in} node accepts incoming HTTP requests. In Figure~\ref{fig-sample-flow}, the
\emph{http in} node is labeled by the HTTP method and the name of the
endpoint, ``{[post]}/webhook''. The \emph{http response} node delivers
the HTTP response. In Figure~\ref{fig-sample-flow}, it is labeled ``http'' and shows up
multiple times as the end of several branches of the flow.

A pair of \emph{rasaas} nodes, \emph{init} and \emph{finish,}
respectively, unpack and store the information carried by an incoming
Rasa action request or combine and transform results of an action into
an HTTP response object coded as JSON according to the Rasa Action
Server API. The \emph{init} node follows the \emph{http in} node. A
\emph{finish} node precedes each \emph{http response} node, as shown in
Figure~\ref{fig-sample-flow}.

\subsubsection{Branching into specific actions}

\begin{figure}     
  \includegraphics[width=0.8\linewidth]{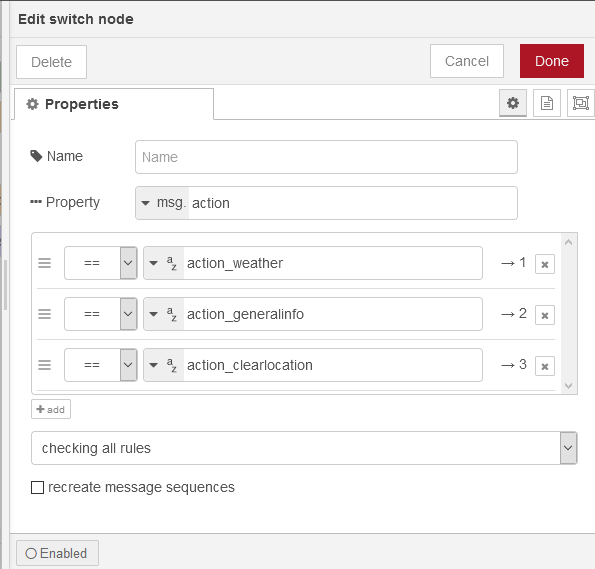}
 % \subfigure[\emph{sendbuttons} node]{\includegraphics[width=0.9\linewidth]{sendbuttons.eps}}
  \caption{Example configuration of a \emph{switch} node in Node-RED}~\label{fig-switch-node-config}
  \Description{ A GUI dialog box with several input fields holding  some text input  }
\end{figure}

The action server in Figure~\ref{fig-sample-flow} implements three different actions, two of
which involve calling an external API (\emph{action\_weather} and
\emph{action\_generalinfo}). A third action,
\emph{action\_clearlocation}, sets a slot value. The chatbot includes
the name of the intended action into its HTTP call to the action server.
The \emph{init} node stores the action name into a specific ``action''
field of the flow's message object, to be evaluated by a \emph{switch}
node. The \emph{switch} node is a standard Node-RED node. In the flow in
Figure~\ref{fig-sample-flow}, it selects the appropriate branch for a particular action. It
has three outgoing branches, each of them corresponding to one of the
three actions. To give an impression of how this actually works, Figure~\ref{fig-switch-node-config} hows the configuration 
view of the \emph{switch} node. The configuration view of a node opens with a double click.

\subsubsection{Generating responses and slot-setting events}

\begin{figure}     
  \includegraphics[width=0.8\linewidth]{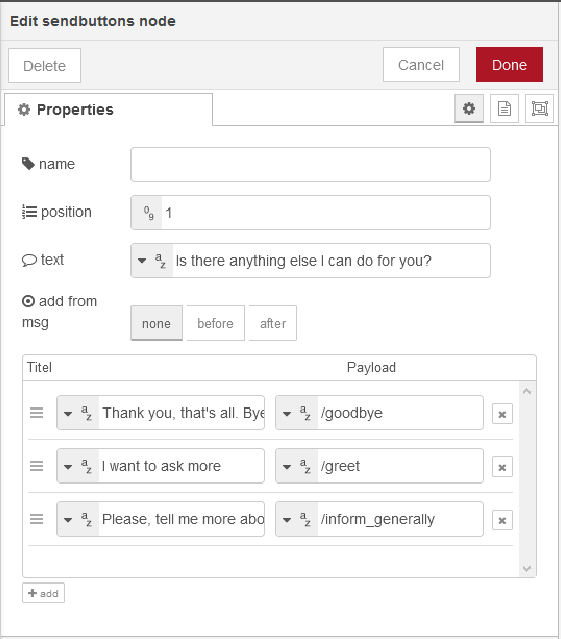}
  \caption{Example configuration of a \emph{sendbuttons} node in Node-RED}~\label{fig-sendbuttons-node-config} 
  \Description{ A GUI dialog box with several input fields holding  some text input   }
\end{figure}

The HTTP response of the action server contains a list of responses for
the chatbot to utter in reply to the user's input, and a list of events
manipulating slots. Responses may take the form of text responses,
attachments, images, buttons, and more. The \emph{rasaas} node
collection provides a node \emph{sendtext} for generating text
responses, a node \emph{sendextra} for generating attachment and image
responses, and a node type \emph{sendbuttons} for generating button
responses. Figure~\ref{fig-sendbuttons-node-config} shows the configuration view of a
\emph{sendbuttons} node. A \emph{setslots} node generates events setting
Rasa slots. A \emph{finish} node collects and assembles the outputs of
these nodes as is shown in Figure~\ref{fig-sample-flow}.

\subsubsection{Calling external APIs}

Calling an external API requires preparing the URL and query parameters
of the API. The results of an API need to be analyzed and transformed
into a chatbot response. The flow in Figure~\ref{fig-sample-flow} uses \emph{template} nodes
for these tasks.

A \emph{template} node is a standard node type in Node-RED. As its name
indicates, the configuration of a \emph{template} node consists of a
string template containing placeholders. When the flow is executed,
values taken from the message object substitute the placeholders.

A \emph{template} node can hold a template URL string with placeholders
for query parameters. The node extracts query parameters from the
incoming message object and inserts them into the URL template. Then, an
\emph{HTTP request} node suffices for executing the API call. Like
\emph{template} nodes, \emph{HTTP request} nodes belong to Node-RED
standard nodes. An exemplary template for an API call is:
\begin{quote}\sloppypar
http://api.weatherstack.com/current?access\_key= xxx \&query=\{\{slots.location\}\} 
\end{quote}

  Double curly braces designate placeholders. The expressions within the
curly braces specify how to extract the substituting values from the
message object.

Another \emph{template} node can hold a response template. When
executed, it inserts parts of the API response result into the
placeholders and propagates the resulting bot response to a
\emph{sendtext} node. An example of such a template string is:
\begin{quote}
It is \{\{payload.current.weather\_descriptions.0\}\} in
\{\{payload. location.name\}\}, \{\{payload.location.country\}\} at the
moment.
\end{quote}

\subsection{Developing, deploying and debugging}

Developing an action server with Node-RED means adding and configuring
nodes and drawing flows. The exemplary action server flow described here
implements three actions. It is easy to implement more actions by adding
more branches to the flow after the \emph{switch} node.

The flow in Figure~\ref{fig-sample-flow} contains one other node type, namely, a standard
Node-RED \emph{debug} node labelled ``msg''. When triggered, a
\emph{debug} node prints the message object or parts of the message
object to a debug tab in the Node-RED GUI, making errors and problems
visible.

The Rasa action server API includes two additional services an action
server should implement: a \emph{/health} service reporting the
operational state of the action server, and an \emph{/actions} service
listing the actions the server is capable to execute. These are easy to
implement by flows starting with corresponding \emph{http in} nodes.

Clicking a Deploy button in the Node-RED GUI immediately activates
modifications or additions to the flow.

This yields very quick and convenient development cycles.

\section{Discussion and future work}

This paper proposes a system architecture in which the open IoT platform
Node-RED serves as middleware for the Fulfillment component of a
conversational UI implemented with the Rasa chatbot development
platform. This approach has several benefits.

From a technical viewpoint, Node-RED is a good fit for a Rasa chatbot,
as both systems are available as open source and deployable on premise,
giving developers and operators full ownership over their systems and
data. The overall architecture is clear and simple, as both systems
remain separate and communicate via HTTP by a compact API. This differs
from the approach of \cite{guidone_guidonenode-red-contrib-chatbot_2020}, which uses Node-RED as development and
deployment platform for chatbots.

The system architecture also differs from chatbot development platforms
like Kore.ai. Kore.ai provides visual programming with flows for
defining dialogs. It also allows calling APIs from within the Kore.ai
dialogs via built-in service nodes \cite{koreai_build_nodate}. In contrast, the idea of the
system architecture proposed here is to apply visual programming for
building a uniform API giving access to a variety of heterogeneous
services. This system architecture resembles the architecture of
combining a Kore.ai chatbot with Blue Prism Robotic Process Automation
via a connector \cite{blue_prism_blue_2019,koreai_botkit_2019,koreai_build_nodate}. 
In the terminology of \cite{baez_chatbot_2020}, these
architectures correspond to the integration pattern \emph{conversational
business process interface}.

The system architecture proposed here is motivated by the intent to
facilitate the development of an action server for the Rasa chatbot to
the point where even non-programmers can successfully build a working
transactional chatbot including its fulfillment component. To date,
experience indicates good success: As an assignment in a lab exercise, a
class of undergraduate students were asked to develop Rasa chatbots with
Node-RED action servers as described in this paper, instructed by a
detailed tutorial, and, in a next step, to extend their bots by
additional and new functions. Despite lacking an IT background or
programming skills, they were able to accomplish this task very
satisfactorily with reasonable effort. In conclusion, Node-RED is a
feasible solution for the problem of no-code or low-code development of
service fulfillment for conversational user interfaces.

The potential of Node-RED goes beyond calling web APIs, since the
community has contributed a wealth of custom nodes that connect to
various systems including databases, spreadsheets, etc. In sum, however,
the ease of programming an interface also depends on the backend
systems. For example, Node-RED has nodes that connect to databases, but
retrieving the correct data still requires knowledge of the database
query language.

So far, the proposed system architecture has been tested in prototyping
and in classroom settings. Next, we intend to explore further
integrations and evaluate the approach for real world applications. 
Rasa chatbot training data and a sample Node-RED flow that demo the approach can
be found on GitHub \cite{weberi_weberirasaas-demo-flow_2021}.

\begin{acks}
The author is thankful to Markus Huber for bringing Node-RED to her attention.
\end{acks}
\bibliographystyle{ACM-Reference-Format}

\end{document}